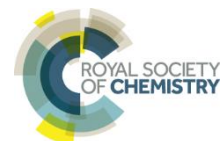

# Journal Name

## ARTICLE

# Enhanced oil removal from water in oil stable emulsions using electrospun nanocomposite fiber mats

S. Barroso-Solares[a,b]*, J. Pinto[a], G. Nanni[a], D. Fragouli[a], and A. Athanassiou[a]*

a Smart Materials, Nanophysics, Istituto Italiano di Tecnologia, via Morego 30, 16163 Genova, Italy.
b Università degli Studi di Genova, via Balbi, 5, 16126, Genova, Italy.
*Corresponding authors: sbarroso@fmc.uva.es, Athanassia.Athanassiou@iit.it

**Abstract**

Fibrous mats with hydrophobic and oleophilic properties have been fabricated and used as absorbents of oil from stable water in oil emulsions. The mats were prepared by initially mixing two polymers, poly(methyl methacrylate) (PMMA) and polycaprolactone (PCL), in a common solvent. The subsequent electrospinning of the prepared solutions resulted in the production of mechanically stable fiber mats, with enhanced oil absorption capacity and oil absorption selectivity from the emulsions, compared to the pure PMMA or PCL mats. Furthermore, the formed fibrous substrates have been successful in the absorption of oil from different emulsions with a wide range of oil content, from 10 to 80 v.%. The performance of the fibrous mats was optimized by the incorporation of hydrophobic silica nanoparticles, reaching oil absorption capacities of 28 g/g and negligible water uptake, in the emulsions with 80 v.% oil content.

## 1.Introduction

Several industrial processes (e.g., in pharmaceutical, cosmetic, and food industries) generate, as waste, water in oil emulsions, with significantly adverse effects on the environment and the human health[1]. Oil in wastewater can be found in various forms, such as free, soluble, and emulsified. Free and soluble oil can be efficiently separated by physical techniques, such as gravity or ultrasonic separation, skimming, absorption, or filtration, with minimal environmental impact and low operating costs[2,3]. Nonetheless, these techniques are not applicable for separation of water/oil emulsions, due to the high stability of the tiny droplets present in the emulsions, which hinders their separation or removal by conventional techniques[4,5]. Recently, emulsified oil wastewaters have been successfully treated using multiple-step demulsification processes, in which the emulsion is first broken inducing flocculation, and then separated using common physical techniques (e.g., absorption)[6,7,8,9,10]. For instance, magnetic nanoparticles (MNPs) were used as chemical demulsifiers, taking advantage of their electrostatic interaction with the oil and their hydrophobicity, to flocculate oil droplets, and then remove the oil attached to the nanoparticles by using their magnetic response[6]. A similar approach was reported by Zang et al.[7], but in this case, the MNPs were coated with silica and 3-aminopropyltriethoxysilane (APTES), to increase the hydrophobic response. Despite the high efficiency of these demulsification processes, a significant drawback should be noted. These approaches present severe risks of producing secondary pollution (i.e. release of the nanoparticles) on the treated/recovered liquids and the environment[11,12].

Trying to avoid these risks, polymeric materials with specific wetting properties (simultaneous hydrophobicity/ oleophilicity)[13,14], have been commonly used as filtration membranes[15,16,17,18,19,20,21,22,23] for the separation of water in oil or oil in water emulsions. These systems still face limitations, such as low efficiency, mainly due to their poor selectivity towards one of the two phases of the emulsions. To improve the selectivity, and therefore the separation performance, porous systems have been developed with enhanced wetting states, i.e. simultaneous superhydrophobicity and superoleophilicity, for the efficient removal of oil from oil in water emulsions, taking advantage of their enhanced oil affinity and increased water repellency[11]. Strategies to achieve these wetting states include exposed surface increase or surface functionalization[11,14,19,22,24,25,26,27,28,29,30]. Among the different methods to fabricate porous structures with appropriate surface structure and composition for optimized wetting properties, the electrospinning method has been considered as one of the most effective[30]. This technique allows the formation of multi-porous structures, with high surface to volume ratio, and specific physicochemical functionalities, provided by the easily controllable composition of the electrospun fibers[31,32]. Nanofibrous mats formed in this way have been recently proposed as sorbents or filters for water-oil separation applications[18,31,32,33,34]. Specifically, the use of polymeric nanofibrous membranes with pronounced wetting properties for the separation of emulsions has been widely studied[18,35,26,27]. Nonetheless, the majority of the so far presented studies is focused on the separation of non-stabilized (surfactant free) emulsions, demonstrating the efficiency of the process by utilizing specific amounts of emulsion, either in the low or in the high oil content range. To the best of our knowledge, there is no demonstration so far that electrospun polymeric fibrous mats can separate stabilized emulsions, in a wide range of concentrations of water in oil.

Regarding the separation of stabilized emulsions, nanocomposite coated electrospun poly(vinyl alcohol) (PVA) nanofiber mats were able to separate by filtration oil in water emulsions with 0.1 wt.% oil concentration, stabilized by a nonionic surfactant[18]. Nanofibrous membranes of polyvinylidenefluoride (PVDF)/stearic acid (SA) were employed to separate efficiently by filtration water in oil emulsions with 98 wt.% oil concentration stabilized using Span80[15]. In another study, nanofiber-assembled cellular *aerogels* produced by freeze-shaping of polyacrylonitrile (PAN) and $SiO_2$ electrospun nanofibers, have been utilized as highly efficient filters for water in oil emulsions stabilized using Span80 as surfactant[30]. These materials exhibit optimal properties, such as superhydrophobic-superoleophilic behavior, but require a complicated fabrication process, while their performance has been demonstrated in emulsions of solely 1.0 wt.% water concentration in oil. To extend the applicability of fibrous mats in a wider range of stabilized emulsions, we recently proposed the use of poly(methyl methacrylate) (PMMA)-based electrospun fiber mats for the separation of oil from stabilized, high internal phase, water in oil emulsions, i.e. with oil contents up to to 30 v.%, reaching an oil absorption capacity of 20 grams of oil per gram of fibers (g/g). Nonetheless, despite their good performance and facile fabrication process, these materials could not be used in water in oil emulsions with oil contents higher than 30 v.%, due to their partial dispersion into the emulsions[34]. In order to extend this emulsion separation approach to a wider range of water in oil emulsions, there is the need to develop electrospun fiber mats with not only the appropriate porosity and wettability but also with the suitable structural integrity and stability in the whole range of oil content in water in oil emulsions.

To do so, we followed a common approach to tune the properties of electrospun fiber mats, which is the solvent blending of two compatible polymers, PMMA and polycaprolactone (PCL). The appropriate selection of the blends' composition plays a key role to the final characteristics of the fiber mats, such as the fibers' morphology and size, fibrous mat's porosity, fibers' entanglement, surface properties, and thermal and mechanical properties[32, 36]. PMMA has often been blended with PCL, a biodegradable, and mechanically robust material,[37, 38, 39, 40] resulting in a composite system with increased thermal and mechanical stability[41, 42, 43]. Therefore, we developed PMMA/PCL electrospun fiber mats, and we present herein their analysis regarding morphology, mechanical and wetting properties and, most importantly, oil absorption capacity and selectivity using a wide range of stabilized water in oil emulsions concentrations. The maximum oil absorption capacity reached is 25.3 g/g for the emulsions with 80 v.% of oil, but with simultaneous absorption of a small amount of water. To further improve the oil absorption capacity of the system, and especially its oil selectivity during the emulsions separation process, we modified the PMMA/PCL fiber mats by introducing 1.0 wt. % of hydrophobic silica nanoparticles as fillers. In this way, we formed highly selective nanocomposite fibrous mats, able to remove oil efficiently from a broad range of water in oil emulsions, with maximum oil absorption capacity 28.0 g/g and negligible water absorption, for the emulsions with 80 v.% oil concentration.

## 2. Materials and Methods

### 2.1. Materials

Poly(methyl methacrylate) (PMMA) ($M_w$ ~120,000 by GPC, $\rho$=1.17 g/cm$^3$ at 25 °C) and polycaprolactone (PCL) ($M_w$ ~80,000 by GPC, $\rho$=1.15 g/cm$^3$ at 25 °C) were obtained from Sigma-Aldrich in the form of pellets. Chloroform ($CHCl_3$, Purity (GC) > 99.8%, $\rho$=1.48 g/cm$^3$ at 20 °C Sigma-Aldrich) was used as received. Hydrophobic silicon oxide fumed nanoparticles (AEROSIL® R 812) with sizes between 5 and 40 nm ($\rho \approx$2 g/cm$^3$ at 20 °C, Evonik Industries AG), were used as nanofillers for the fabrication of the nanocomposite fibers. Light mineral oil ($\rho$=0.84 g/cm$^3$ at 25 °C, Sigma-Aldrich) and distilled water were mixed to obtain emulsions, using Span80 (Sorbitan monooleate, non-ionic, viscosity 1200-2000 mPa·s at 20 °C, HLB value 4.3±1.0, $\rho$=0.986 g/cm$^3$ at 25 °C, Sigma-Aldrich) as an emulsion stabilizer.

<u>Preparation PMMA/PCL fiber mats:</u> First, PCL and PMMA were dissolved separately in $CHCl_3$ at concentrations of 0.164 g/mL and 0.370 g/mL, respectively, by shaking for 24h at room temperature (RT).

Then, the PCL and PMMA solutions were mixed in the appropriate proportions to obtain PMMA-PCL blends with weight ratios of 30/70, 50/50 and 70/30 (PMMA-PCL) and were shaken for 24h at 800 rpm and RT[42]. The solutions employed for the nanocomposite fibers' fabrication were obtained by adding 1.0 wt.% (concerning the total weight of the polymer blend) of AEROSIL® R 812 nanoparticles (SNPs) into the 50/50 PMMA-PCL solution. The prepared PCL, PMMA, and PMMA-PCL (with and without SNPs) solutions were charged in a syringe (diameter of 4.5 mm and 21G needle) which was then placed into a syringe pump (NE-1000, New Era Pump Systems, Inc.) of a fixed flow rate of 1000 μl/h. A grounded target covered with aluminum foil was used as a collector, and the distance between the needle tip and the collector was 20 cm, whereas the voltage was set at 12 kV. Under these conditions, pure PCL fibers (PCL), pure PMMA fibers (PMMA), PMMA_PCL (at 30/70, 50/50 and 70/30 weight ratios) fibers (PMMA/PCL), and 50/50 PMMA/PCL nanocomposite fibers with 1.0 wt.% of SNPs (SNP/PMMA/PCL) were fabricated. All the fibers were left under a fume hood for 12h to ensure the complete evaporation of the $CHCl_3$. As shown in our previous work,[34] the neat PMMA fibers require a press post-processing before their utilization in water in oil emulsions separation due to their poor structural integrity. Therefore, before their utilization in the emulsions separation experiments and wetting characterization, the PMMA fibers were compacted using a hot plate press, kept at 50 °C and 98 kPa for 5 min[34]. For the mechanical and morphological characterization, non-compacted PMMA fibers were utilized in order to carry out a proper comparison with the other fiber mats, which were not subjected to the press process due to their superior performance.

## 2.2. Characterization of the fiber mats

The fiber mats morphology was examined with scanning electron microscopy (SEM, JEOL Model JSM-6490). Energy Dispersive Spectroscopy (EDS) was employed to identify the presence and distribution of the SNPs on the surface of the fibers. The fibers were coated with a thin gold layer (10 nm) using a Cressington 208HR sputter coater (Cressington Scientific Instrument Ltd., U.K.) prior to the SEM observations. Their average diameter and distribution were determined using FIJI / ImageJ[44].

The porosity of the fiber mats (also known as volume fraction of voids ($V_f$)) is defined as the ratio between the gaseous phase volume and the total volume of the porous sample. It was calculated according to the Equation 1, where $\rho_r$ represents the relative density of the porous material, and equals to the ratio between the bulk density of the fiber mats ($\rho_f$) and the density of the corresponding solid materials ($\rho_s$)[34] [45] $\rho_f$ is determined by the weight divided by the geometrical volume of the fiber mats, while the $\rho_s$ of PCL, PMMA, 70/30, 50/50, 30/70 PMMA-PCL blends, and of the 50/50 PMMA-PCL blend with 1.0 wt.% of nanoparticles, which can be obtained by the rule of mixtures, are 1.145, 1.170, 1.153, 1.158, 1.630, and 1.170 g/cm³, respectively.

$$V_f = 1 - \rho_r = 1 - \rho_f/\rho_s \qquad [1]$$

The mechanical properties of the fiber mats were experimentally determined by uniaxial tensile tests using an Instron 3365 machine. Tensile testing was conducted under environmentally controlled conditions (20 °C, 40% RH). To collect stress-strain curves, the samples were stretched at a rate of 5 mm min$^{-1}$. The Young's modulus, ultimate tensile strength (UTS), and elongation at maximum load were extracted from the resulting curves. Five specimens for each type of sample were tested (PCL, PMMA, as well as 70/30, 50/50, and 30/70 PMMA/PCL, and SNP/PMMA/PCL with SNPs 1.0 wt.%), and the results were averaged to obtain a mean value and the standard deviation in each case.

Finally, the water contact angle (WCA) and oil contact angle (OCA) average values and standard deviations were determined by taking 5 measurements for each kind of sample using 5 μl drops, with a KSVCAM200 (Kruss, Germany) contact angle goniometer at RT.

### 2.3. Preparation and characterization of the emulsions

To from stable water in oil emulsions using light mineral oil, one surfactant with low Hydrophilic-Lipophilic Balance (HLB) values, between 4 and 6, is needed. For this reason, the Span80 (HLB value 4.3±1.0) was chosen. The emulsions were prepared by dissolving Span80 into the oil phase (0.5 wt.%) and shaking for 24h[46]. Then, the aqueous phase was added to the mixture and sonicated to form the Span80-stabilized emulsions, using a high intensity ultrasonication tip (VCX 750, Vibra cell, SONICS) at 40% amplitude for 15 secs at RT[34]. The concentrations of the emulsions are expressed using the volume percentage of the oil (v.%) in water. In general, for each absorption experiment, 2 ml of water in oil emulsions were prepared with oil contents of 10-30-50 and 80 v.%, which correspond to oil amounts of 0.168-0.504-0.840-1.344 g, respectively.

In order to control the stability over time (5, 15, 30, and 60 min) of all the water in oil emulsions (10, 30, 50, and 80 v.% of oil) a Nikon 139 Eclipse 80i Digital Microscope was employed for their morphological characterization. To prepare the samples for the microscope analysis, a small amount of the emulsion (one drop) was placed on a microscope glass slide. Then it was covered with a coverslip, and finger pressed to make it as thin as possible. Finally, the drop size distribution was analyzed using FIJI / ImageJ[44] on the obtained micrographs for all the aforementioned time intervals.

### 2.4. Absorption capacity experiments

The absorption capacity ($C$) experiments were performed using three samples of each type of fiber mats. Samples of 0.02 g and dimensions of ~2.000x2.000x0.025 cm$^3$ were placed into 2 ml of pure water, pure oil, or emulsions. The maximum oil or water absorption capacity ($C_{max}$) of the fiber mats was calculated by measuring the weight of the fibers before ($w_b$) dipping into pure oil or pure water for 15 minutes, and immediately after ($w_a$) their extraction from the liquids (Equation 2)[47]. To investigate the performance of the developed materials for the emulsions separation and calculate the oil absorption capacity ($C_{oil}$), the samples were dipped in the water in oil emulsions for 15 minutes under shaking. To this aim, the weight of the fibers was measured before ($w_b$) and immediately after ($w_a$) their removal from the emulsions, and again after being stored for 1 week at 50º C ($w_{a'}$) to ensure the complete water evaporation from the samples. The oil absorption capacity was calculated by dividing the remaining liquid (oil) absorbed weight ($w_{a'}$ - $w_b$) after 1 week, to the initial weight of the fibers ($w_b$) (Equation 3); whereas the simultaneous water absorption capacity ($C_{water}$) was determined by Equation 4.

$$C_{max} = (w_a - w_b)/w_b \qquad [2]$$
$$C_{oil} = (w_{a'} - w_b)/w_b \qquad [3]$$
$$C_{water} = (w_a - w_{a'})/w_b \qquad [4]$$

## 3.Results and Discussion

Among the different kinds of fiber mats produced by electrospinning (PCL, PMMA, and PMMA/PCL), the PMMA ones were the less compact, and thus, more difficult to handle. Indeed, the PCL and PMMA/PCL fibers were forming macroscopically homogeneous and compact membranes during the electrospinning process, indicating that the presence of PCL improves the structural integrity of the fiber mats, forming durable substrates as shown in Fig. 1a. In Fig. 1b-f, where are shown SEM images of the different electrospun mats, it is clear that the morphology and diameter distribution of the PMMA fibers (Fig. 1b) are affected by the addition of the PCL. Although in all cases bead-free randomly distributed fibers are formed, the 70/30 and 50/50 PMMA/PCL (Fig. 1d, 1e) have smaller mean diameters (c.a. 3.1±1.0 and 3.4±1.1 μm, respectively) with a narrower distribution (ranging between 1 and 6 μm) compared to the PMMA fibers (mean diameter 6.5±1.9 μm and range 2-12 μm) and to the PCL fibers (mean diameter 4.6±1.2 μm and range 2-7 μm) (Fig. 1b and c respectively). On the other hand, the diameters of the 30/70 PMMA/PCL fibers (Fig. 1f) have a pretty broad and bimodal distribution, with a population of fibers in the same range of sizes as the other PMMA/PCL fibers, and a second population of fibers with diameters below the micron. This indicates that for the specific combination different electrospinning parameters should be explored in order to obtain homogeneous fibers.

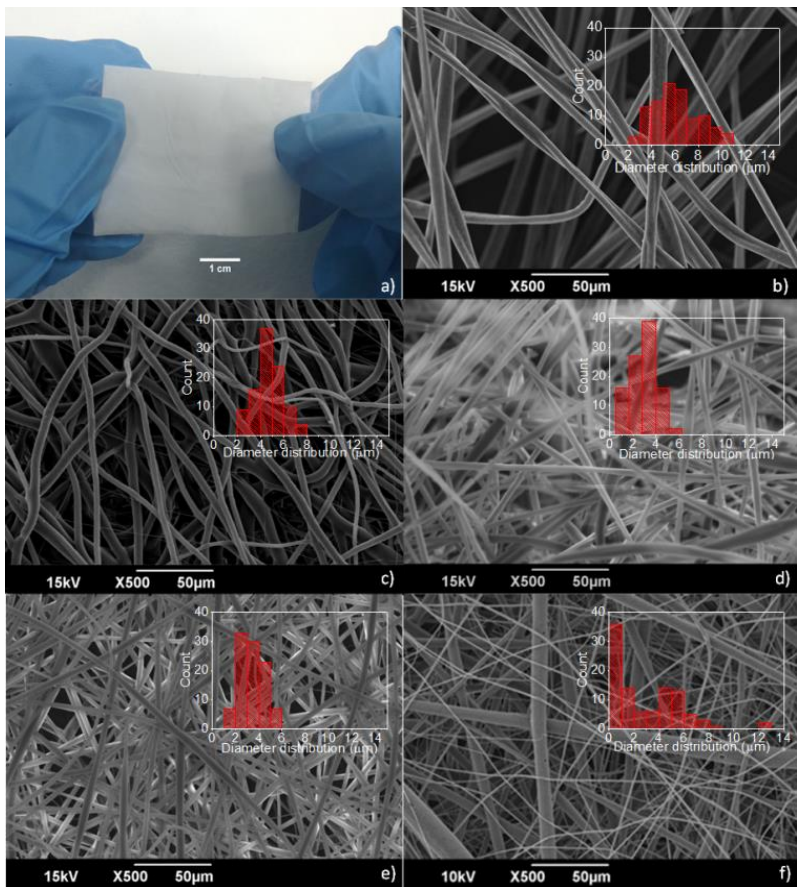

Fig. 1. (a) Photograph of the fibrous mat 50/50 PMMA/PCL, and SEM micrographs of fibrous mats: (b) PMMA, (c) PCL (d) 70/30 PMMA/PCL, (e) 50/50 PMMA/PCL, and (f) 30/70 PMMA/PCL. Insets: the diameter distributions of the fibers in each case.

In order to evaluate the effect of the PCL polymer on the structural integrity of the fabricated fibers, the mechanical properties of the different mats are studied and summarized in Table 1.

Table 1. Mechanical properties of PMMAFbs and PCPMFbs.

| Fiber mats | PMMA (wt.%) | PCL (wt.%) | Young's modulus (MPa) | UTS (MPa) | Elongation at maximum load (%) |
|---|---|---|---|---|---|
| PMMA | 100 | 0 | 6.4±3.2 | 0.03±0.01 | 3.8±1.3 |
| 70/30 PMMA/PCL | 70 | 30 | 11.2±3.9 | 0.15±0.03 | 2.2±0.3 |
| 50/50 PMMA/PCL | 50 | 50 | 18.5±2.7 | 0.26±0.03 | 24.0±1.4 |
| 30/70 PMMA/PCL | 30 | 70 | 20.9±2.5 | 1±0.2 | 31.7±7.9 |
| PCL | 0 | 100 | 13.6±1.9 | 1.10±0.1 | 77.7±23.3 |

Pure PMMA fiber mats have Young's modulus of 6.42±3.23 MPa. When PMMA is combined with PCL, the Young's modulus of the correspondent PMMA/PCL fibers increases with the amount of PCL in the composition, indicating a higher resistance of the fibers to deformation. Similarly, the ultimate tensile strength (UTS) and the elongation at maximum load are positively affected by the presence of PCL, with a maximum increment with respect to the PMMA mats of 3060% and 735%, respectively, for the 30/70 PMMA/PCL combination (additional details about the mechanical characterization can be found in the Supporting Information S.I-1). It should be noted that 50/50 and 30/70 PMMA/PCL fibers present even higher Young's modulus than the PCL ones, demonstrating a good interaction between the two blended polymers. Furthermore, as the mechanical properties of the fiber mats depend not only on the polymer properties but also on the structure of the mats, this is also a good indicator of the synergy between the PMMA and PCL polymers for the production of highly entangled electrospun fibers.

To investigate the possibility to use the composite fibers as oil adsorbents for water-oil separation, their wettability properties were first determined, as this is a crucial factor for the materials' ability to separate emulsions[48]. The WCA of pure PCL and PMMA fibers was respectively 138°±2° and 130°±2°. Accordingly, the hydrophobicity of PMMA/PCL fibers increased by increasing the PCL content, with WCA values of 135°±3° for the 70/30 PMMA/PCL fibers, 142°±1° for the 50/50 PMMA/PCL fibers, and 145°±4° for the 30/70 PMMA/PCL fibers, with the values of the last two being even higher than the ones of the pure PCLFbs. This is most likely attributed to the increased surface roughness of these two kinds of samples, resulting in an increased area of air (WCA=180°) trapped under the measured water drops, and thus in augmented WCAs[28, 29]. The increased roughness in the case of the 50/50 PMMA/PCL fibers can be due to the smaller diameter of the composite fibers with respect to the PCL fibers and, in the case of the 30/70 PMMA/PCL, to the inhomogeneous diameter distribution of the fibers[49].

However, despite their high WCA values, the fibers' surfaces in all cases present a significant water adhesion, which can lead to some water entrapment between the membranes' voids upon their immersion in water. This is confirmed by dipping the fibers for 15 minutes in distilled water and measuring their water absorption capacity, using equation (2). The results are presented in Fig. 2. The PMMA fibers have a water absorption capacity of 2.0±0.2 g/g, slightly higher than the capacity of PMMA/PCL ones (1.8±0.2 g/g). Unexpectedly, the 70/30 and 50/50 PMMA/PCL fibers present higher water absorption capacities than the neat polymer fibers, with 4.00±0.50 g/g and 2.5±1.6 g/g, respectively. On the other hand, the 30/70 PMMA/PCL mats have water absorption capacity of the same range as the pristine polymers (1.6±1.6 g/g).

Regarding the oil wettability of the produced fiber mats, when oil droplets are placed on the mats' surface, they rapidly spread and penetrate the structure within few seconds (OCA≈0º), independently of the fibers' composition. After dipping the fiber mats in pure oil, the maximum oil absorption capacity is determined using equation (2), and Fig. 2 shows its dependence on the mats' composition. In the case of the pure polymers, the PCL fibers present lower oil absorption capacity (3.9±0.3 g/g) than the PMMA fibers (18.0±0.3 g/g), and therefore it was expected that the combination of PCL with PMMA could deteriorate the oil absorption capacity of the PMMA fibers. Nevertheless, as shown in the Fig. 2, the maximum oil absorption capacity of the PMMA/PCL fibers increases with the PCL concentration, with values of 23.0±3.6 g/g for 70/30 and 25.0±1.4 g/g for 50/50 of PMMA/PCL mats. Mats with higher PCL contents (i.e., 30/70 PMMA/PCL) show a significant decrease of the absorption capacity, which reaches 10.0±0.5 g/g. Having in mind that the porosity of the developed mats is similar in all cases (see Supporting Information S.I-2) we can conclude that there is no relationship between the porosity of the dry fiber mats and their oil absorption capacity, as oil absorption capacities range from 3.9 to 25.0 g/g for fiber mats presenting the same porosity. We assume that this behavior is related to the modifications of the three-dimensional structure of the fiber mats upon interaction with the oil during the absorption tests. In particular, the porous structure of the PCL-rich fiber mats is not changing its conformation upon interaction with oil, presenting thus low oil absorption capacities. On the contrary, when the PMMA component becomes important in the composition of the mats, their overall volume significantly increases upon oil adsorption, resulting in higher oil absorption capacities (see Supporting Information S.I-2). In these cases, the PCL component is responsible for the conservation of the structural integrity of the mats. In fact, the low structural integrity of the pure PMMA fibers upon oil absorption is the limiting factor for their oil absorption capacity.

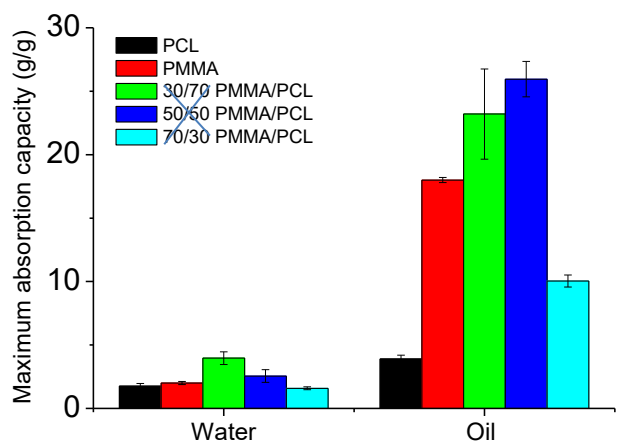

Fig. 2. Maximum water and oil absorption capacity of the fibers after 15 min of full immersion in the respective liquids.

Subsequently, the performance of the developed fibrous systems as oil absorbers in stable water in oil emulsions is presented. Initially, different stable water in oil emulsions were prepared (10, 30, 50, 80 v.% of oil) using Span80 as emulsifier (see Supporting Information S.I-3), with the vast majority of droplets' diameter well below 2 μm. As shown in Fig. 3a, although the oil absorption capacities (measured as described in the experimental part and calculated using equation 3) of PMMA/PCL fiber mats from emulsions with 10 and 30 v.% oil, are lower compared to PMMA fiber mats, it is noteworthy that using the PMMA/PCL mats it is possible to perform demulsification experiments to emulsions of oil concentrations up to 80 v.%. In fact, PMMA mats reach their maximum oil absorption capacity (19.8±0.5 g/g) in 30 v.% oil emulsions, and this is the emulsion with the maximum oil content in which these fibers can be used without losing their structural integrity, being dispersed in the emulsion.

On the other hand, PCL fiber mats are mechanically stable upon immersion in all emulsions but show an almost constant and low oil absorption capacity (about 3.9±0.3 g/g) independently of the oil concentration in the emulsion. However, when mixed with PMMA, their performance changes completely with the 70/30 and 50/50 PMMA/PCL mats to reach maximum oil absorption capacities of 25.3±1.1 g/g and 24.4±0.2 g/g, respectively, for emulsions with the highest oil content (80 v.%). The 30/70 PMMA/PCL are much less efficient to remove the oil from the emulsions, with oil absorption capacities below the one of the PMMA, independently of the oil content, possibly attributed to the high amount of PCL, which reduces their performance.

Regarding the water absorption simultaneously with the oil during the emulsions separation experiments (Fig. 3b), PCL and PMMA fiber mats present rather low water absorption capacities (calculated using equation 4), while PMMA/PCL mats show a similar behavior like in the pure water absorption tests. In particular, for the 70/30 PMMA/PCL fiber mats the water absorption capacity from the emulsions is about 9-10 g/g regardless of the emulsion, and for the 50/50 PMMA/PCL fibers, it ranges from 10.86±3.20 g/g in 10 v.%-oil emulsions to about 3.2±1.9 g/g in 80 v.%-oil emulsions. The 30/70 PMMA/PCL fibers present a negligible water absorption from the emulsions, due to both their higher hydrophobicity and to their low total absorption capacity.

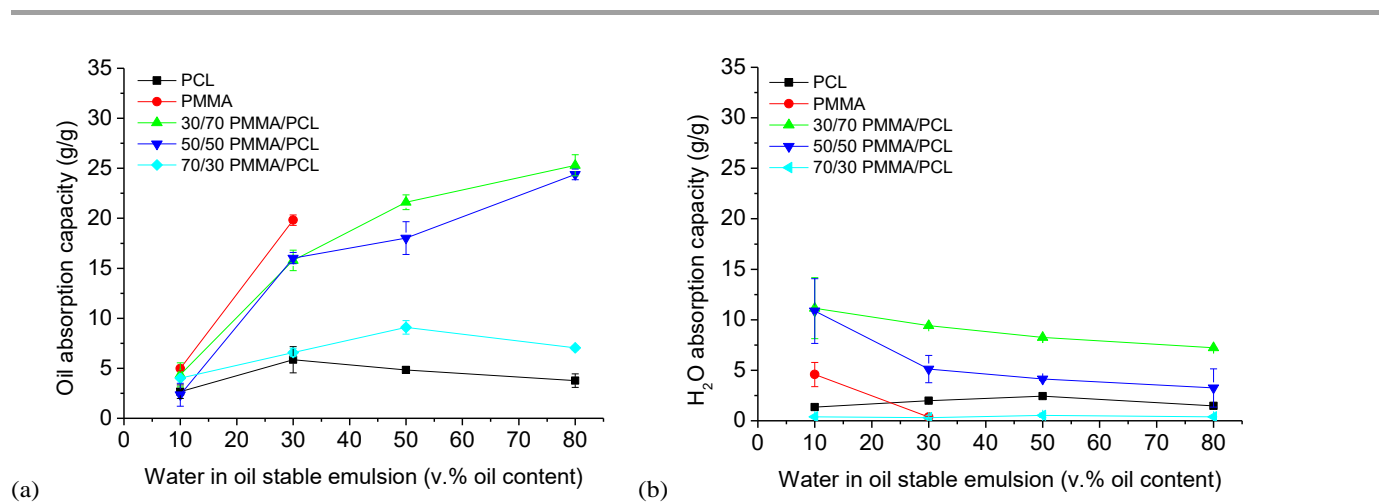

Fig. 3. Oil absorption capacity (a) and Water absorption capacity (b) of PCL, PMMA, and PMMA/PCL fiber mats in emulsions with oil contents from 10 to 80 v.%.

Therefore, despite their good oil absorption performance, both from emulsions and pure oil (Fig. 2), the PMMA/PCL fiber mats do not achieve optimal oil selectivity in the emulsions during the water-oil separation process, since they simultaneously absorb water. For this reason, a further modification of the PMMA/PCL mats is necessary in order to reach the desired selectivity during the separation process, by increasing the hydrophobicity of the system. Since 50/50 PMMA/PCL mats show the most promising compromise between the oil absorption capacity, oil selectivity, and mechanical properties, this blend was selected for optimization of its performances, by adding SNPs in the fibers' structure.

Specifically, PMMA/PCL nanocomposite fibers (SNP/PMMA/PCL) were obtained by adding 1.0 wt.% of SNPs to the 50/50 PMMA/PCL fibers with the aim to improve their oil selectivity without modifying their desirable features, such as the high oil absorption capacity and applicability in the entire range of emulsions under study. As shown in Fig. 4a, the nanocomposite fibers are bead-free and have a smooth surface, presenting the same diameter range as the 50/50 PMMA/PCL fibers (i.e. 1-6 μm) and a comparable average diameter (3.0±1.0 μm for the SNP/PMMA/PCL). Similarly, the porosity is in the same range as the one of the mats without the SNPs, as shown in the Supporting Information Table S.I-1. In addition, the presence of the SNPs on the surface of the SNP/PMMA/PCL was proved by EDS characterization, showing a clear signal from the Si, quite uniformly distributed on the surface of the fibers (Fig. 4b).

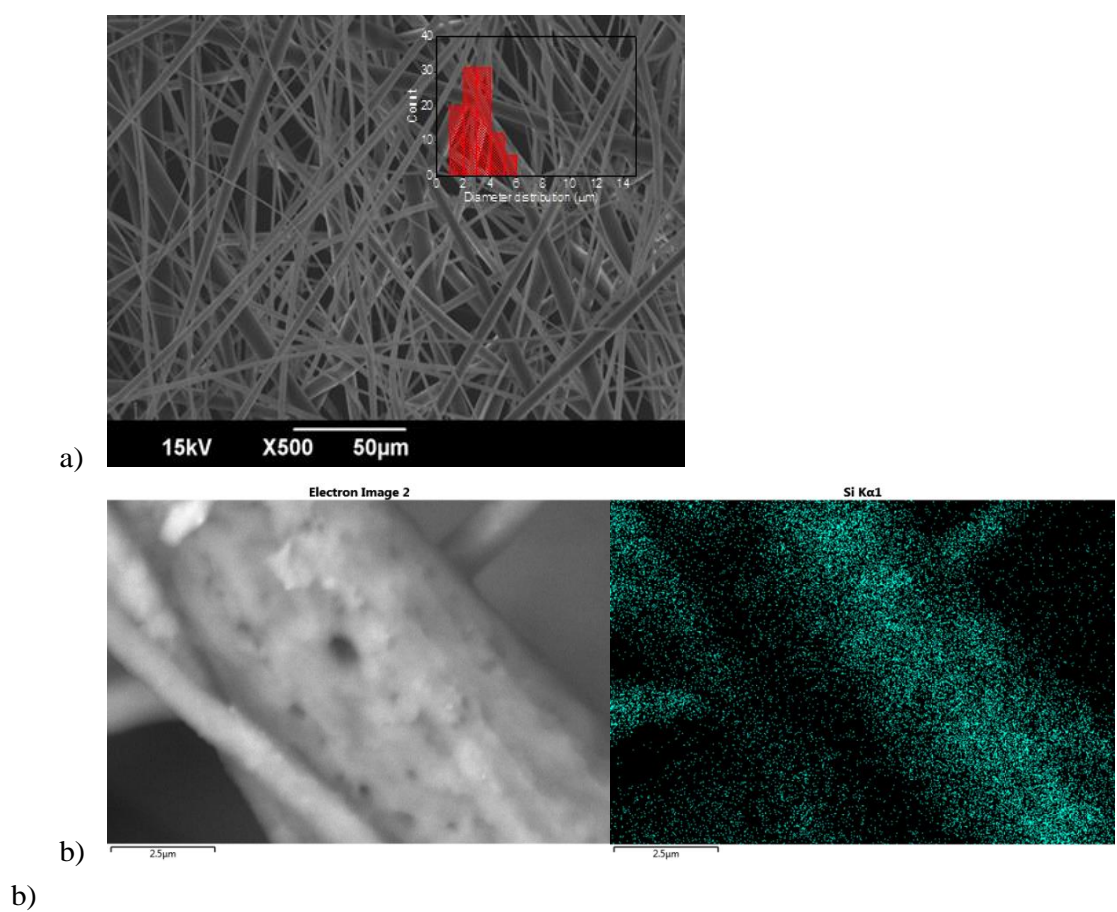

a)

b)

b)

Fig. 4. a) SEM micrographs and diameter distribution (inset) of SNP/ PMMA/PCL fibers with 1.0 wt.% SNPs, and b) SEM image with higher magnification and the corresponding EDS map showing a homogeneous distribution of Si (green color) on the surface of the fibers.

Although the morphologies of the nanocomposite and the polymer blend fibers are similar, the mechanical properties of the PMMA/PCL fiber mats change after the incorporation of the nanoparticles (see Supporting Information S.I-4), yet being appropriate for their use in emulsions with high oil contents. The addition of SNPs improves the hydrophobicity of the mats, by increasing the WCA by 5° (150°±1°) with respect to the original 50/50 PMMA/PCL fibers, while no significant changes were found regarding the oleophilicity of the fibers. This enhancement of the hydrophobicity led to a significant decrease of the maximum water absorption capacity when tested in pure water (calculated using equation 2), reaching values of 1.55±0.50 g/g instead of 2.5±1.6 g/g of 50/50 PMMA/PCL fibers. At the same time, the nanocomposite mats have an increased maximum oil absorption capacity, when tested in pure oil, of 28.4±3.1 g/g (calculated using equation 2). Therefore, it is confirmed that the addition of SNPs improves the selectivity of the fiber mats towards oil absorption, by both increasing the oil absorption capacity and decreasing the water absorption.

Fig. 5 shows a comparison between the oil and water absorption capacity of 50/50 PMMA/PCL fiber mats and their nanocomposites SNP/PMMA/PCL mats from water in oil emulsions with oil contents from 10 to 80 v.%. The enhancement of the oil selectivity from the fiber mats after the incorporation of SNPs, presented above, is responsible for remarkable results in the absorption tests from stable water in oil emulsions. Although the SNP/PMMA/PCL fiber mats show a similar oil absorption with 50/50 PMMA/PCL mats in water in oil emulsions with oil contents up to 50 v.%, they reach a higher oil absorption in water in oil emulsions with 80 v.% of oil. Most importantly, the SNP/PMMA/PCL mats present a remarkably improved behavior in comparison to the 50/50 PMMA/PCL mats in terms of simultaneous water absorption, with the nanocomposite fibers presenting practically no water uptake in the entire range of emulsions (Fig. 5).

Therefore, it is established that the incorporation of SNPs in the 50/50 PMMA/PCL fibers (SNP/PMMA/PCL mats) is necessary, in order to obtain a suitable material for water-oil separation by oil absorption from stable water in oil emulsions. The SNP/PMMA/PCL mats reach optimal oil selectivity in the emulsions' separation tests (i.e. negligible simultaneous water uptake), and as enhanced maximum oil absorption capacity of 28.4 g/g (for oil content 80 v.% in the emulsions), indicating that the fibrous nanocomposite mats are ideal for the proposed application reaching a high efficiency of separation. Although it is not possible a direct comparison of the performance of these fibers as oil absorbents with previous works from the literature, in which fiber mats have been successfully employed as filters for stable emulsions[15, 30], it could be highlighted that the developed fibers not only reach an optimal oil selectivity, but also keep that selectivity for a wide range of emulsions (i.e., a wide range of oil contents from 10 to 80 wt.%), while previous works in the literature only considered one fixed oil content.

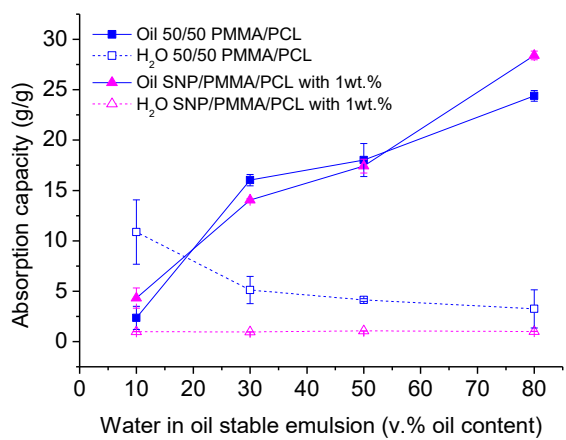

Fig. 5. Oil absorption capacity and Water absorption capacity of 50/50 PMMA/PCL and SNP/PMMA/PCL fiber mats in emulsions with oil contents from 10 to 80 v.%.

## 4.Conclusions

In summary, we demonstrate that electrospun fiber mats composed by PMMA and PCL can effectively remove oil from stable water in oil emulsions in a large range of oil contents. In particular, combining the right concentrations of PMMA and PCL, the fiber mats can have significantly higher oil absorption capacity than the fiber mats made of each polymer separately. The proposed PMMA/PCL fiber mats can be efficiently used in a wide range of emulsions with oil contents from 10 to 80 v.%. The ability to remove oil from high oil content emulsions without any damage of the fiber mats was found to be related to the mechanical properties and the good entanglement of the PMMA/PCL fibers. Addition of 1.0 wt.% of hydrophobic silica nanoparticles in the 50/50 PMMA/PCL fibers, results in an optimization of the oil selectivity and of the oil absorption capacity of the system, reaching the final value of oil absorption of 28 g/g with a simultaneous negligible water absorption from all the stable emulsions tested. The possibility to form the fiber mats with a straightforward and single step process and to use them efficiently in a wide range of stabilized emulsions makes them a suitable candidate for the remediation of the oil present in emulsified industrial wastewaters.

## 5.Aknowledgements

The authors acknowledge Simone Lauciello (Electron Microscopy Facility, Istituto Italiano di Tecnologia) for his assistance with the HRSEM analysis.

## 6.References